\documentclass[iop,apj,numberedappendix]{emulateapj}
\usepackage{amsmath}
\usepackage{txfonts}
\usepackage{bm}
\usepackage{cancel}

\def\d{{\rm d}}
\def\rs{r_{\rm s}}
\def\rc{r_{\rm c}}
\def\rh{r_{\rm h}}
\def\Rh{R_{\rm h}}
\def\Rhr{R_{{\rm h,r}}}
\def\Rhp{R_{{\rm h,p}}}
\def\rhoz{\rho_{0}}

\def\Klosone{{K_{{\rm los,p}}}}
\def\Klostwo{{K_{{\rm los,r}}}}
\def\Wlosone{{W_{{\rm los,p}}}}
\def\Wlostwo{{W_{{\rm los,r}}}}
\def\sigmar{{\sigma_{\rm r}}}
\def\sigmap{{\sigma_{\rm p}}}
\def\mur{{\mu_{\rm r}}}
\def\mup{{\mu_{\rm p}}}
\def\nur{{\nu_{\rm r}}}
\def\nup{{\nu_{\rm p}}}
\def\ar{{a_{\rm r}}}
\def\ap{{a_{\rm p}}}

\begin{document}

\journalinfo{\sc
prepared to be submitted to
\sl the Astrophysical Journal}

\shorttitle{A Virial Core in Sculptor}
\title{A Virial Core in the Sculptor Dwarf Spheroidal Galaxy}
\shortauthors{\sc Agnello and Evans}
\author{A. Agnello and N. W. Evans}
\affil{Institute of Astronomy, University of Cambridge, Madingley Road, Cambridge CB3 0HA, UK}
\email{aagnello@ast.cam.ac.uk (AA), nwe@ast.cam.ac.uk (NWE)}

\begin{abstract}
The projected virial theorem is applied to the case of multiple
stellar populations in the nearby dwarf spheroidal galaxies. As each
population must reside in the same gravitational potential, this
provides strong constraints on the nature of the dark matter halo. We
derive necessary conditions for two populations with Plummer or
exponential surface brightnesses to reside in a cusped
Navarro-Frenk-White (NFW) halo. We apply our methods to the Sculptor
dwarf spheroidal, and show that there is no NFW halo compatible with
the energetics of the two populations. The dark halo must possess a
core radius of $\sim 120$ pc for the virial solutions for the two
populations to be consistent. This conclusion remains true, even if
the effects of flattening or self-gravity of the stellar populations
are included.
\end{abstract} 

\section{Introduction}

Using high resolution numerical simulations of structure formation,
Navarro, Frenk \& White (1995) first proposed that dark haloes have
the universal form
\begin{equation}
 \rho(r) = {\rhoz \over (r/\rs)(1 + r/\rs)^2}
 \label{eq:nfwdef}
\end{equation}
where $r_s$ is the scale-radius and $\rhoz$ is the density
normalisation. Sometimes it is useful to define a halo radius $\rh$,
which is the radius within which the mean density is a factor of
$\Delta_{\rm h} = 200$ times greater than the critical density. Then,
the enclosed mass of the NFW profile is
\begin{equation}
  M(r) = 4 \pi \rhoz \rs^3 \Biggl[ \log (1 + cx) - {cx \over 1 + cx}
    \Biggr]
\end{equation}
where $x = r/\rh$ and $c = \rh/\rs$. The two parameters of the NFW
halo can be alternatively chosen as the concentration $c$ and the mass
$M = M(\rh)$.  The NFW model remains the best two-parameter fit to
simulation data, though \citet{Na04} provided evidence that the
three-parameter Einasto profile gives a still better match.

Despite an extensive body of research, it is unclear whether the
shapes of dark haloes inferred from observations are consistent with
cosmological predictions. In galaxies like the Milky Way and M31, the
dark halo profiles have been significantly affected by the cooling of
baryons, which makes them unattractive candidates for testing the
prediction. Therefore, research has concentrated on two very different
classes of objects, namely galaxy clusters and dwarf galaxies. In
clusters, the high temperature and low density of the intra-cluster
gas prevents efficient cooling, and so the baryons do not significantly
affect the potential of the dark matter. There is reasonably good
evidence from cluster lensing that the profiles follow the NFW law,
although with higher concentrations than predicted (e.g., Broadhurst
et al. 2008).

In dwarf galaxies at the present epoch, the baryonic component is
feeble in comparison with the dark matter.  Low surface brightness
galaxies with HI and H$\alpha$ rotation curves have been studied
extensively by \citet{Bl02}, who conclude that cored rather than
cusped haloes are favoured.  The dwarf spheroidal galaxies have no
gas, and are pressure-supported rather than rotation-supported, so are
more difficult to assess. It was quickly realised that the radial
velocities of thousands of bright giant stars could be gathered with
modern multi-object spectrographs~\citep{Kl02,Wa07}. Dynamical
modelling with either the Jeans equations or phase space distribution
functions then yields constraints on the dark matter profile. However,
this line of inquiry seemed to have stalled with the realization of
the degeneracies inherent in the Jeans equations, making the datasets
compatible with both cored and cusped dark halo models depending on
the underlying assumptions~\citep{Ev09,Wa09}.

The problem has received new impetus with the growing realization that
dwarf spheroidals often contain two or more stellar
populations~\citep{Ba06,Ko06}. This significantly strengthens any
constraints from dynamical modelling, as now each sub-population must
separately be in equilibrium in the same potential. There have been
three recent and thought-provoking results. First, \citet{Ba08} used
the Jeans equations to model the velocity dispersion profiles of the
two populations in the Sculptor dSph, finding that a cored halo is
preferred to an NFW halo, though the latter is still statistically
consistent with the observations. Second, using the same dataset but
more rigorous methods based on distribution functions, \citet{Am11}
showed that NFW halos are rejected at high significance, performing
substantially poorer than cored models. Finally, \citet{Wa11} used the
half-light radii of the two populations in the Fornax and Sculptor
dSphs to estimate the enclosed mass. This exploited earlier results
showing that the mass within the half-light radius is robust against
changes in the anisotropy~\citep{Wa09, Wo10}. Given the enclosed mass
at two radii, the slope of density is inferred and shown to be in
contradiction with NFW haloes.

Here, we provide a simple line of reasoning based on the energies of
the subpopulations. This brings the earlier arguments of \citet{Ba08},
\citet{Wa11} and \citet{Am11} into sharp focus. We show that the
energetics of the subpopulations in one of the best studied dSphs,
Sculptor, cannot be made consistent with an NFW halo.

\begin{figure*}
	\centering
        \includegraphics[width=0.3\textwidth]{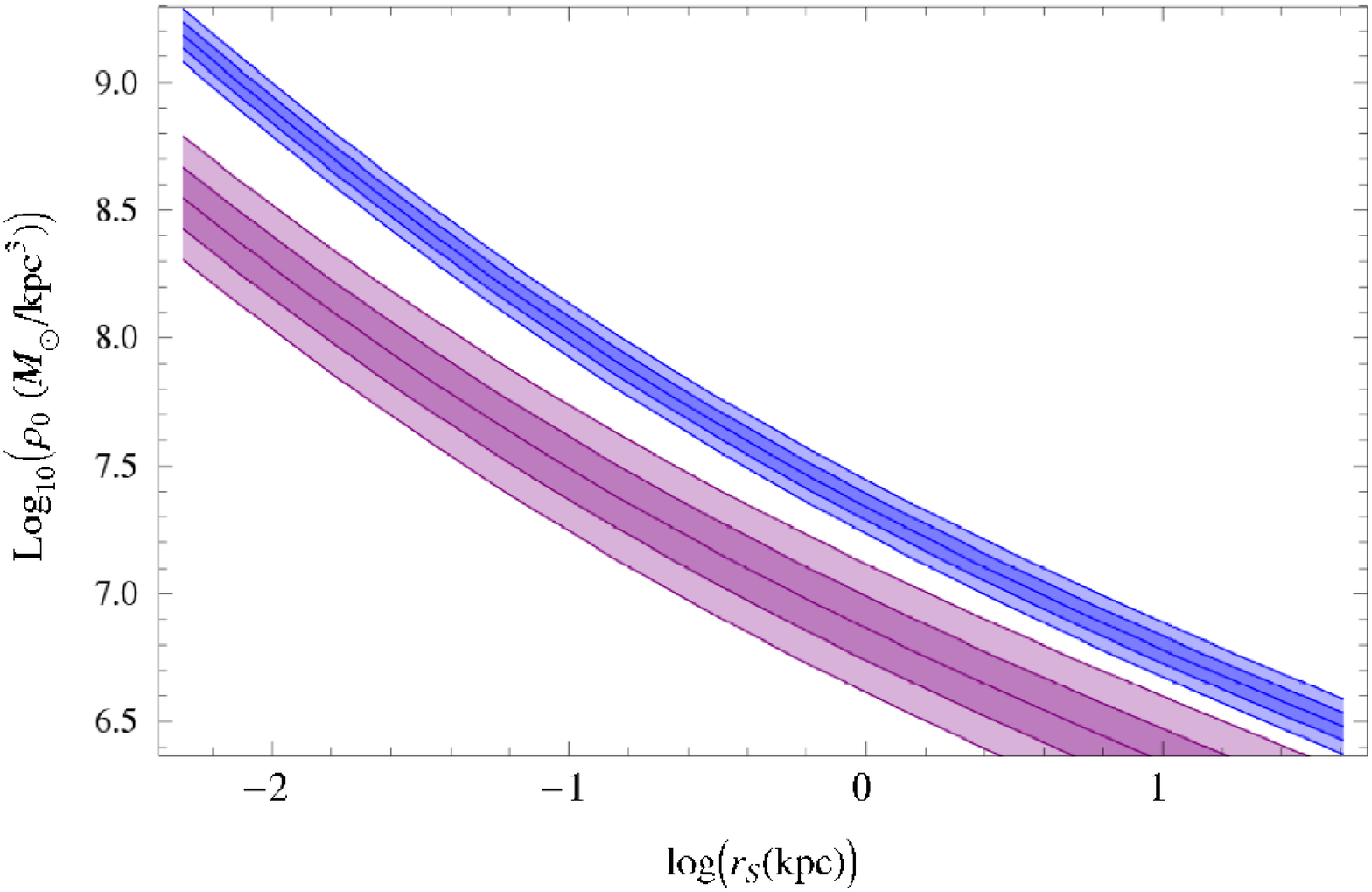}
\includegraphics[width=0.3\textwidth]{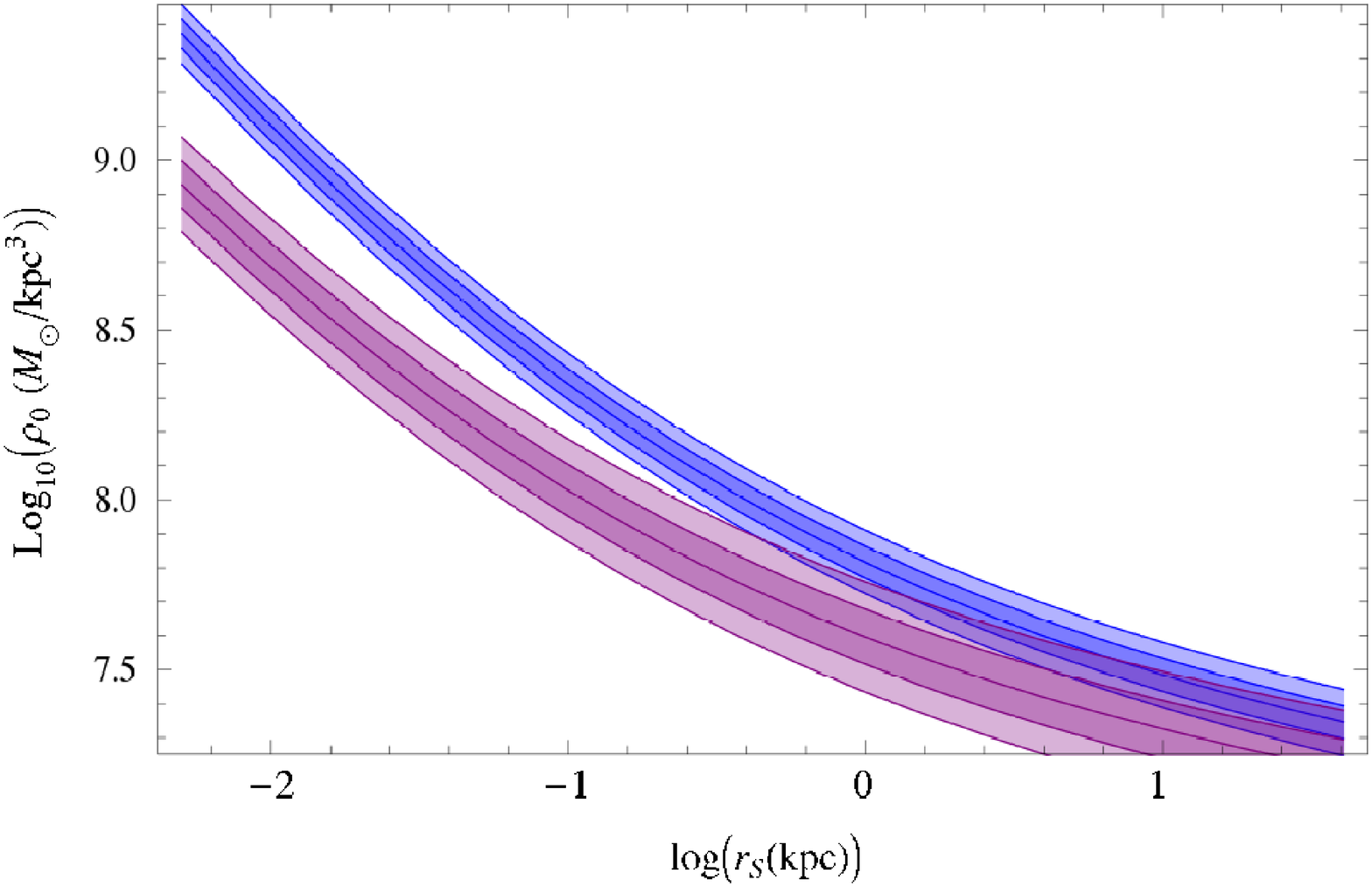}
\includegraphics[width=0.3\textwidth]{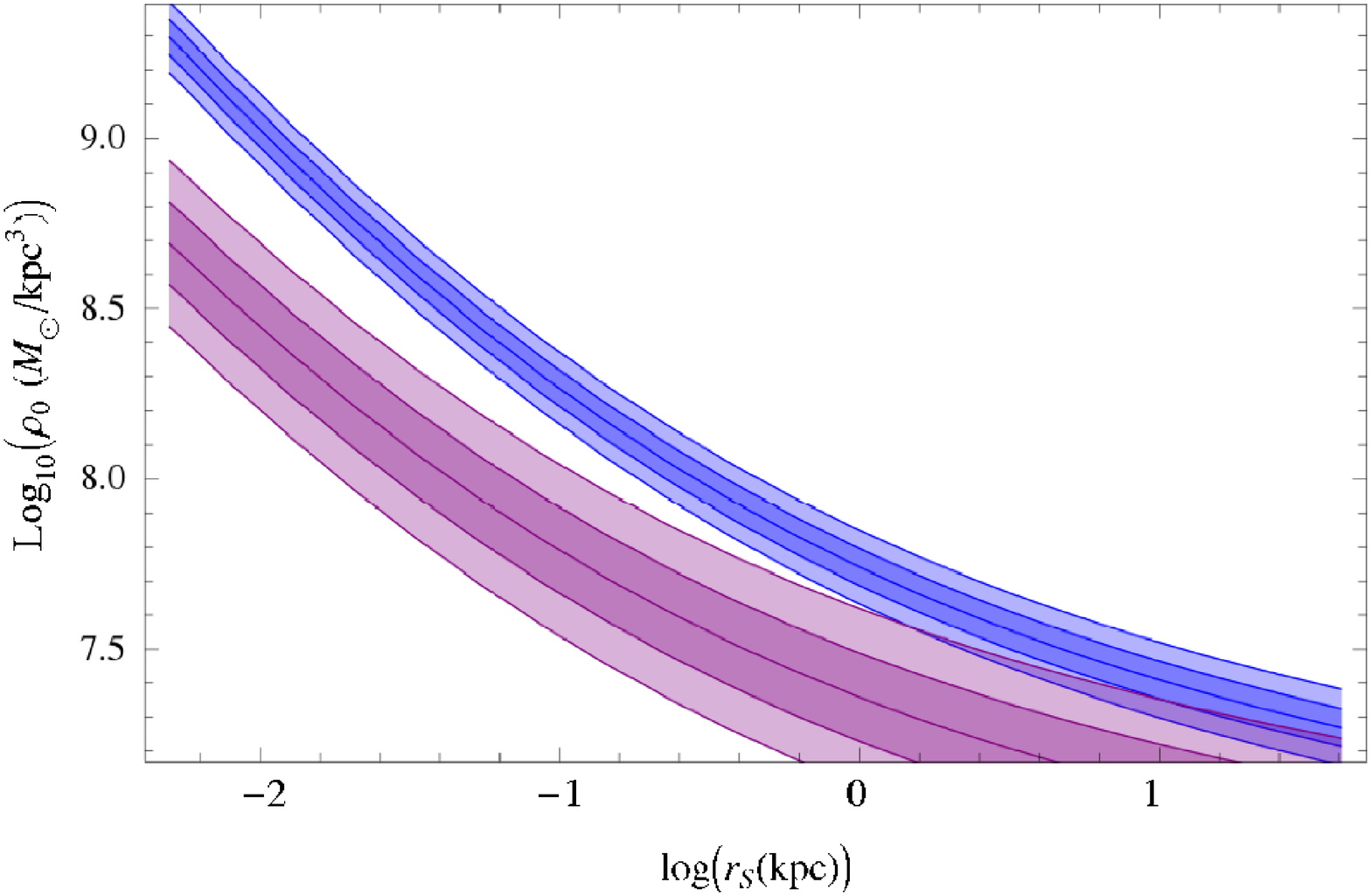}
\caption{\small{ Left: Virial stripes for the two stellar populations
    in Sculptor in a cusped NFW potential, including the self-gravity
    of the stellar populations ($\Upsilon_\star =8$).  Purple shows
    the metal-rich population, blue the metal-poor population. In each
    stripe, the central line is the mean value of
    $\log_{10}(\rho_{0}),$ whilst the median and outer lines follow
    the 1$\sigma$ and 2$\sigma$ deviations. Center and Right: Virial
    stripes for the two stellar populations in a NFW potential with
    small core, without ($\Upsilon_\star =0$) and with ($\Upsilon_\star
    =8$) self-gravity.}}
\label{fig:vscored}
\end{figure*}

\section{The Projected Virial Theorem}

Let $z$ be the line of sight direction. Then, the projected virial
theorem relates the projected component of the pressure and potential
energy tensors by the equation $2K_{\rm los} + W_{\rm los} =0$, where
\begin{equation}
K_{\rm los} = {\Upsilon_\star\over2} \int \nu \langle v_{\rm z}^2\rangle
\d^3x,\qquad\qquad W_{\rm los} = -\int \nu z \partial_{z} \Phi \d^3x,
\end{equation}
Here, $\nu$ is the luminosity density of a population with stellar
mass-to-light ration $\Upsilon_\star$ moving in the gravity field
$\Phi$. Under the assumption of spherical symmetry, the tensors
become~\citep[e.g.,][]{Me90}
\begin{eqnarray}
K_{\rm los} &=& \pi\Upsilon_\star \int_0^\infty \mu \langle v_{\rm
  los}^2 \rangle R \d R \\ W_{\rm los} &=& -{4 \pi G \over
  3}\Upsilon_\star \int_0^\infty r\nu(r) M(r) \d r,
\end{eqnarray}
where $M(r)$ is the mass enclosed within radius $r$.  These formulae
involve the line of sight velocity second moment $\langle v_{\rm
  los}^2 \rangle$ and the surface brightness $\mu$, which are directly
accessible to observation. Notice that the velocity anisotropy of the
stellar population does not enter, which gives the virial equations an
immediate advantage over the Jeans equations. Notice also that virial
quantities, which are gross volume integrals, can be computed more
robustly than the gradients of observables required by the Jeans
equations.  Furthermore, there is no problem in dealing with rotation,
as the integral $K_{\rm los}$ requires just the sum of square
velocities in ordered and random motions along the line of sight.

The only unknown in the projected virial equation is the total
gravitational potential $\Phi(r)$ in which the population moves, or
equivalently the enclosed mass $M(r)$. If preferred, $W_{\rm los}$ can
be recast in terms of the total mass density $\rho(r)$ which generates
$\Phi$, namely
\begin{equation}
W_{\rm los} = -{16 \pi G \over 3}\Upsilon_\star \int_0^\infty R \mu(R)
\int_0^R \rho(r) {r^2 \d r \over \sqrt{ R^2 - r^2}} \d r \d R.
\end{equation}
Here, we have eliminated the luminosity density in terms of the
surface brightness, which is an observable.

The projected virial equation by itself is of course insufficient to
determine the potential or mass distribution directly.  Suppose now
there are two populations in equilibrium in the same total
potential~\citep[e.g.,][]{Ag12}. In the dSphs, a younger metal-rich
and an older metal-poor population with associated velocity dispersion
profiles $\sigmar$ and $\sigmap$ and surface brightnesses $\mur$
and $\mup$ (with corresponding luminosity densities $\nur$ and
$\nup$) are often present. For example, there is good evidence that
Sculptor~\citep{Ba08,Wa11}, Carina~\citep{Bo10} and
Fornax~\citep{Am12b} contain at least two distinct populations. Then,
the projected virial theorem must be satisfied for each population, so
we have
\begin{equation}
{\Klosone \over \Klostwo}  =  {\Wlosone \over \Wlostwo} \implies
{\int_0^\infty \mup \sigmap^2 dR \over \int_0^\infty \mur \sigmar^2
  dR} =
{\int_0^\infty \nup M(r) r\d r.\over \int_0^\infty \nur M(r) r \d r.}
\label{eq:ratios}
\end{equation}
The only unknown is the potential, or equivalently, the enclosed mass
$M(r)$. This result is powerful enough to rule out some dark matter
haloes. The physical reason is that a stellar population with velocity
dispersion ${\langle v^2_{\rm los} \rangle}$ offers the best
constraint on the potential at radii given by $\Phi(r) \approx \langle
v^2_{\rm los} \rangle$. With more than one populations, the potential
is restricted at more than one location.

\section{Application to Sculptor}

\subsection{Simple Results}

If the potential is heavily dominated by dark matter, we can neglect
the self-gravity of the luminous components in a first approximation
(relaxed in the next sub-section). In this case, both the kinetic
energy tensor $K_{\rm los}$ and the potential energy tensor $W_{\rm
  los}$ scale linearly with the stellar mass-to-light ratio, which can
then be eliminated.

The surface brightness profiles of dSphs are well-fit by Plummer laws
\begin{equation}
\mu(R) = {\mu_0 \over (1 + R^2/\Rh^2)^2} \qquad .
\label{eq:plum}
\end{equation}
This assumption is not critical to our argument, but it does provide a
useful illustrative model. We shall in any case sketch the extension
to another commonly used profile, namely the exponential law. Fitting
the Plummer law to the profiles of the metal-rich and metal-poor
populations in Sculptor gives $\Rhr = 230 \pm 10$ pc for the
metal-rich and and $\Rhp = 350 \pm 10$ pc for the metal-poor
  (consistent with the values in \citet{Ba08} and \citet{Am11}). A
  convenient fitting formula for the velocity dispersion profile is
\begin{equation}
{\langle v^2_{\rm los} \rangle} = {\sigma_0^2 \over (1 + R^2/a^2)}\ ,
\label{eq:vdisp}
\end{equation}
which gives a flattish profile with amplitude $\sigma_0$ out to a
characteristic lengthscale $a$. This is typical of the velocity
dispersion profiles in, for example, Walker et al. (2007) or Battaglia
et al. (2008). Again, this assumption is not necessary for our
argument, as the relevant virial integrals can always be computed
numerically. For Sculptor, fits provide $\sigma_{0,{\rm r}} = 8.7 \pm
1.0$ kms$^{-1}$ and $\ar = 240 \pm 55$ pc for the metal-rich and
$\sigma_{0,{\rm p}} = 10.9 \pm 0.8$ kms$^{-1}$ and $\ap = 1920 \pm
850$ pc for the metal-poor.

With these laws, the kinetic energy integral is analytic with
\begin{equation}
K_{\rm los} = {\pi\over 2}
\Upsilon_{\star}\mu_0 \sigma_0^2 a^2\frac{a^{2}/\Rh^{2}\
  -1\ -2\log(a/\Rh)}{(1-a^{2}/\Rh^{2})^{2}}\ .
\end{equation}
The metal-rich limit is given by $\ar \approx \Rhr$, the metal-poor
limit is given by $\ap \gg \Rhp$. This gives
\begin{equation}
{\Klosone \over \Klostwo} = 2 
\left({\mu_{0,{\rm p}} \over \mu_{0,{\rm r}}} \right)
\left( {\Rhp \over \Rhr} \right)^2
\left ( {\sigma_{0,{\rm p}} \over \sigma_{0,{\rm r}}} \right)^2\ .
\label{eq:simple}
\end{equation}
The projected potential energy of Plummer profiles embedded in NFW
halos is also exact, as
\begin{eqnarray}
W_{\rm los} &=& -{8\pi^2 G \Upsilon_\star \mu_0 \rhoz \Rh^3 \rs^3 \over 3\Delta^5} \left[
  \Delta(\rs^2 + 3\rs\Rh - 2\Rh^2) \noindent \right.\nonumber  \\ &+&
  \left. \Rh(\Rh^2-2\rs^2) \log \left( {\Rh(\Delta+\Rh) \over \rs
    (\Delta-\rs)} \right) \right]\ ,
\end{eqnarray}
with $\Delta^2 = \Rh^2 + \rs^2$. With the values of $R_{\rm h}$ listed
above, the virial ratio $W_{\mathrm{los,p}}/W_{\mathrm{los,r}}$ as a
function of $\rs$ is increasing and bound from above. In particular,
we have that
\begin{equation}
{\Wlosone \over \Wlostwo} < \left({\mu_{0,{\rm p}} \over \mu_{0,{\rm r}}} \right)
\left( {\Rhp \over \Rhr} \right)^3.
\label{eq:asymptotic}
\end{equation}
Hence, a necessary condition for a NFW halo to support two stellar
populations with Plummer profiles is
\begin{equation}
\left ( {\sigma_{0,{\rm r}} \over \sigma_{0,{\rm p}}} \right)^2 > 2 \left( {\Rhr
  \over \Rhp} \right) .
\end{equation}
This is identical to eq (22) of Amorisco \& Evans (2012a), derived
under different assumptions. If, instead of Plummer profiles,
exponential laws are used to fit the surface brightness profiles, then
the numerical factor becomes $1.9$ instead of $2$ in
eqn~(\ref{eq:simple}). The analogue of eqn~(\ref{eq:asymptotic}) is
unchanged, so that the necessary condition for an NFW halo to support
two stellar populations with exponential surface brightness profiles
is
\begin{equation}
\left ( {\sigma_{0,{\rm r}} \over \sigma_{0,{\rm p}}} \right)^2 > 1.9\left( {\Rhr
  \over \Rhp} \right)\ .
\end{equation}
Using the best-fitting values provided above for Sculptor, it is
immediate to check that the NFW potential is ruled out. Note that the
constraints are simply the requirement that there is a NFW model with
$\rs<\infty$. This is a much looser constraint than requiring
consistency with an NFW model with a concentration $c \approx 20$, as
predicted by cold dark matter theories.

\begin{table}
\begin{center}
\renewcommand{\tabcolsep}{0.2cm}
\renewcommand{\arraystretch}{0.5}
\begin{tabular}{| l | c | c | c |}
    \hline 
$\epsilon = \rc/\rs$ & $\rs$ [in kpc]& $\rs$ [in kpc]& $\rs$ [in kpc]\\  
$\null$ & ($\Upsilon_{\star}\rightarrow0$) & ($\Upsilon_{\star}=4$) & ($\Upsilon_{\star}=8$) \\  
    \hline
1 & 0.72 & 1.06 & 1.23 \\
0.5 & 0.94 & 1.40 & 1.54 \\
0.25 & 1.2 & 1.92 & 2.20 \\
0.125 & 1.6 & 2.88 & 3.28 \\
0.0625 & 2.4 & 4.48 & 4.96 \\
    \hline
  \end{tabular}
\caption{Minimum $\rs$ for two-sigma overlapping of the virial
stripes.}
\label{tab:Sculptor}
\end{center}
\end{table}

\subsection{The Virial Stripes}

The simple results already suggest that the energetics of the two
populations are inconsistent with an NFW profile. However, it is
prudent to confirm this result numerically, discarding some of the
simplifying assumptions made above.
 
Since the measured profiles come with errors, we operate in the
following manner. For each value of $\rs$, we compute $\rho_0$
separately for the two populations for many different
photometric~(\ref{eq:plum}) and kinematic~(\ref{eq:vdisp})
profiles. We weight each result with the likelihood of the fit. Then,
varying $\rs$ produces a \textit{virial stripe} for each population in
the ($\rho_0,\rs$)-plane.  If the two stripes overlap at $2 \sigma$ at
a particular $\rs$, then the model for the potential is plausible at
the $2 \sigma$ level.  Nothing prevents us from including the
contribution of the luminous tracers to the potential as well. The
virial equations then depend also on the stellar mass-to-light ratio
$\Upsilon_{\star}$, which may be different for the two
populations. The projected potential energy $W_{\rm los}$ has a
contribution $W_{\mathrm{dm}}$ from the dark component and a
correction $W_{self}$ from the two luminous ones. For Plummer
profiles, we have for the $i-$th population:
\begin{equation}
W_{self,i}=\pi^2
G\mu_{0,i}R_{h,i}\sum\limits_{j}\Upsilon_{\star,j}\mu_{0,j}R_{h,j}^{2}w\left(
R_{{\rm h},i}/ R_{{\rm h},j}\right)\ ,
\end{equation}
with
\begin{equation}
w(x)=\frac{x^{3}\left[(5x^{2}+3)\mathrm{K}(1-x^{2})-(x^{2}+7)\mathrm{E}(1-x^{2})\right]}{3(1-x^{2})^{3}}\ ,
\end{equation}
where $\rm K,\rm E$ are complete elliptic integrals and
$\Upsilon_{\star,j}$ is the luminous mass-to-light ratio of the $j-$th
population.  As the $\mu_{0,j}$ are given by number counts and not
directly by luminosities, a common rescaling is applied to both
populations such that the total luminosity is fixed at the observed
value (taken from table 6 in Irwin \& Hatzidimitriou 1995).

The leftmost panel of Figure~\ref{fig:vscored} shows the virial
stripes for the two populations in Sculptor, excluding and including
the effects of the self-gravity for the luminous component. The two
stripes never overlap at the $2\sigma$ level. This confirms the result
deduced from our simple argument in the previous section: there is no
NFW halo compatible with the kinetic energies of the two stellar
components in Sculptor.  The center and rightmost panels of
Figure~\ref{fig:vscored} show the virial stripes when the dark halo
density is the simplest cored analogue of the NFW halo, namely
\begin{equation}
\mathrm{cNFW}=\frac{\rhoz}{(\epsilon^{2}+r^{2}/\rs^{2})^{1/2}(1+r^{2}/\rs^{2})}\ .
\end{equation}
In this case, the stripes do overlap at the 2$\sigma$ level provided
the core radius $r_{\rm c}\equiv r_{\rm s}\epsilon$ is at least $150$
pc, if the self-gravity of the stellar populations is neglected.
Incorporating self-gravity causes the core radius to increase
somewhat, as we see in Table~\ref{tab:Sculptor}. This shows how the
minimum $\rs$ for 2$\sigma$ overlap varies with changing
$\Upsilon_{\star}$ for different models. The first column
($\Upsilon_{\star}\rightarrow 0$) stands for models in which the
self-gravity of the stars is omitted. Since both half-light radii are
smaller or equal to $\rs$ in the dark matter only case, adding
self-gravity is expected to yield larger cores, as in fact is
confirmed by the results in the Table.

\begin{figure}
	\centering
        \includegraphics[width=0.4\textwidth]{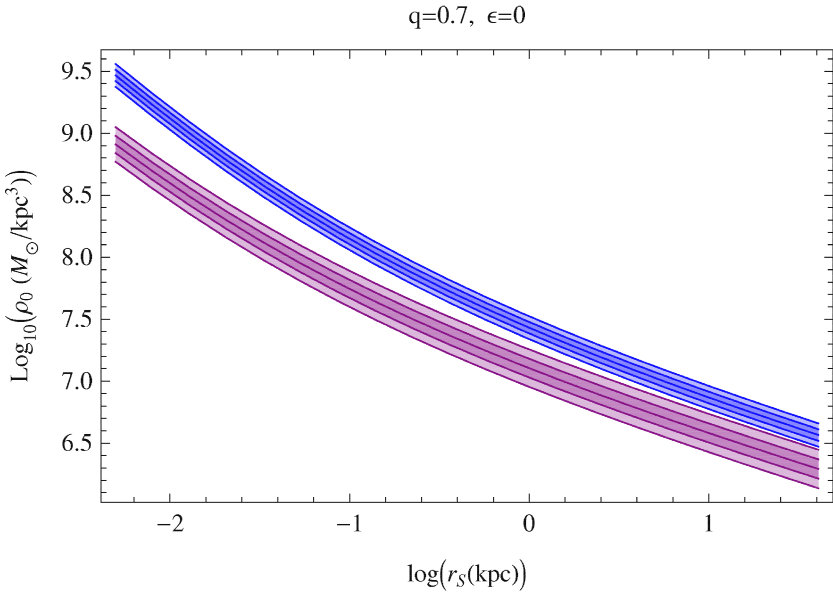}
\caption{\small{Virial stripes for the two stellar populations in
    Sculptor in a flattened NFW potential ($q/g = 0.7$ with
    $\Upsilon_\star =0$).  Purple shows the metal-rich population,
    blue the metal-poor population. In each stripe, the central line
    is the mean value of $\log_{10}(\rho_{0}),$ whilst the median and
    outer lines follow the 1$\sigma$ and 2$\sigma$ deviations.}}
\label{fig:flat}
\end{figure}

\section{Discussion and Conclusions}

The arguments in this {\it Letter} show that the kinematics of
multiple populations in dSphs provide a substantial challenge to the
predictions of cold dark matter cosmogonies. In the case of one of the
best studied dSphs, Sculptor, there is no NFW dark halo that is
compatible with the available photometric and kinematics data.  The
problem is very basic -- the energetics of the metal-rich and
metal-poor populations do not permit them to reside in the same NFW
halo. Are there are any loop-holes?

\subsection{Reliability of Separation of Sub-Populations ?}

This {\it Letter} has used the separation proposed by ~\citet{Ba08},
who used a hard cut in metallicity to define the metal-poor ([Fe/H] $<
-1.7$) and metal-rich populations ([Fe/H] $> -1.5$). This is open to
the objection that the kinematics and metallicity are interlocked, and
so the cut may be subject to unperceived biases.  The sub-populations
can also be separated using a maximum likelihood approach with
metallicity and kinematics treated jointly. We have checked our
calculations using such a method on the sample of stars in
\citet{To04} (also used by Battaglia et al. 2008).  \citet{Wa11} have
also tacked this problem, using a Bayesian likelihood estimator based
on positional, kinematical and chemical data to separate the
sub-populations. Their algorithm also makes assumptions that may or
may not be warranted (for example, the velocity distributions are
assumed isothermal). Nonetheless, although the details of the
separation are different, \citet{Wa11} come to the same conclusion
that the multiple populations of Sculptor are inconsistent with an NFW
halo. Different methods of separation, using different algorithms,
come to the same answer, which offers comfort to those concerned about
reliability.

\subsection{Steady state, well mixed populations ?}

An assumption in the virial theorem is that the stellar populations
are in a steady-state and well-mixed. Recently, de Boer et al (2012)
have provided a star-formation history for Sculptor. There was a peak
in star formation $\sim 13-14$ Gyr ago, providing the bulk of the old,
metal-poor population. Star formation continued till $\tau_{0}\sim 7$
Gyr ago. With the typical dark matter densities found from the virial
stripes, the dynamical time-scale $(4\pi G\rho_{0})^{-1/2}$ is at
least one order of magnitude smaller than $\tau_{0}$. It seems that
the two populations have indeed reached a steady-state and that each
of them is internally well-mixed.  From the {\it Hubble Space
  Telescope} proper motions, the orbital period of Sculptor is
estimated as $\sim 2.2$ Gyr ~\citep{Pi06}. Its pericentric distance is
$\sim 68$ kpc, and the time of the last pericentric passage was $\sim
1$ Gyr ago. As the dSph always remains in the outer parts of the
Milky Way halo, it seems very improbable that tidal stirring can
overturn the steady-state hypothesis.

\subsection{Flattening ?}

A drawback of all previous models of dSphs is that they are restricted
to spherical symmetry. This is not the case for our virial arguments.

The isophotes of the Sculptor dSph have a measured ellipticity of
$0.3$~\citep{Ir85}. In fact, the flattening of the dark matter
potential must be rounder than the flattening of the luminosity
density, so spherical models are probably a good approximation. As a
check, we generalize our virial arguments to flattened stellar
populations (with axis ratio $q$) in a flattened NFW dark matter halo
(with axis ratio $g$).  Let us assume that Sculptor is viewed
edge-on,and the density and kinematics of both stellar populations are
stratified on surfaces $x^2 + y^2 q^{-2}$. We also make the natural
assumption that the minor axes of the dark matter and the luminous
matter are aligned.  Then, each kinetic energy term in
eqn~(\ref{eq:ratios}) scales by $q$, and so the ratio is left
unchanged. Similarly, each potential energy term in
eqn~(\ref{eq:ratios}) is modulated by the same function of $q/g$ in
the integrand (c.f., the Flattening Theorem of Agnello \& Evans 2012)
and so the ratio is left largely unchanged. This suggests that our
virial arguments will be robust against flattening.

To confirm this, we carried out explicit calculations of the virial
stripes for flattened models of Sculptor with $q/g = 0.7$, an example
of which is shown in Fig~\ref{fig:flat}. The virial stripes for the
flattened NFW halo never overlap. Introducing a small core does permit
the kinematics of the two populations to be consistent at the
$2\sigma$ level providing the core is $\gtrsim 120$ pc, a touch
smaller than in the spherical case.

\medskip
We believe that the virial arguments are particularly robust and
emphasise the important point that the energetics of the two
populations in Sculptor are completely incompatible with a $1/r$ dark
matter density cusp. For at least this dwarf galaxy, we can securely
state that it either never had a $1/r$ density cusp, or if it did have
one, feedback processes have now removed it. 

Of course, there has been much recent interest in converting cusped
haloes into cored. A number of promising possibilities have been
suggested, including impulsive mass loss from supernovae~\citep{Re05},
winds and gas flows driven by supernovae~\citep{Ma06} and infalling
baryonic clumps~\citep{Co11}. Which of these processes, if any, were
responsible for the core in Sculptor is an important question for
future work.

\acknowledgments AA thanks the Science and Technology Facility Council
and the Isaac Newton Trust for the award of a studentship. We wish to
thank Nicola C. Amorisco and the anonymous referee for critical
readings of the manuscript, and Giuseppina Battaglia and Mike Irwin for
generously providing us with data.

\clearpage

\end{document}